\documentclass[aps,pra,showpacs,showkeys,twocolumn]{revtex4}
\usepackage{hyperref}
\mathchardef\ordinarycolon\mathcode`\:     
\mathcode`\:=\string"8000
\def\vcentcolon{\mathrel{\mathop\ordinarycolon}}
\begingroup \catcode`\:=\active
  \lowercase{\endgroup
  \let :\vcentcolon
  }

\newcommand{\ket}[1]{|#1\rangle}
\newcommand{\ketbra}[2]{|#1\rangle\!\langle#2|}

\newcommand{\smfrac}[2]{\mbox{$\frac{#1}{#2}$}}

\begin{document}

\title{Comment on ``Quantum identification schemes with entanglements''}
\author{Wim van Dam}
\email{vandam@cs.berkeley.edu}
\altaffiliation{
Computer Science Department, 
University of California, Soda Hall, Berkeley, CA
94720-1776, USA.
Also at Mathematical Sciences Research Institute, 
Berkeley, and HP Labs, Palo Alto. 
This work was made possible by an HP-MSRI 
postdoctoral fellowship.}
\affiliation{UC Berkeley, MSRI, HP Labs Palo Alto}

\date{\today}

\begin{abstract}
In a recent paper, [Phys.\ Rev.\ A \textbf{65}, 052326 (2002)], Mihara
presented several cryptographic protocols that were claimed to be
quantum mechanical in nature.  In this comment it is pointed out that
these protocols can be described in purely classical terms.  Hence,
the security of these schemes does not rely on the usage of
entanglement or any other quantum mechanical property.
\end{abstract}

\pacs{03.67.Dd} 
\keywords{quantum cryptography, classical cryptography, entanglement}
\maketitle The cryptographic protocols presented in \cite{mihara} use
entangled quantum states and are called ``quantum
password schemes'', ``quantum identification schemes'' and ``quantum
message authentication schemes'', thus suggesting that the security of
these procedures relies on the properties typical of quantum information.  
In this Comment we will show that this suggestion is incorrect as these
protocols can be described in purely classical terms.  More
specifically, the protocols in \cite{mihara} only use
the classical correlations between the bits of the distributed
quantum states.  All procedures are, in fact, only variations of the
traditional \emph{Vernam cipher} scheme, which has been known since
the early 1900s \cite{vernam}.
 
The two-party ``quantum password'' scheme of \cite{mihara} uses
$n$ copies of the bipartite, entangled state $\ket{\psi_i}:=
(\ket{0,0}+\ket{1,1})/\sqrt{2}$.  The password is an $n$-bit string
$p\in\{0,1\}^n$, and the goal of the scheme is to let one party (the
system) verify that the other party (the user) knows $p$,
under the requirement that this verification does not make $p$ public.  
This is achieved by the following steps.
First, the system selects a random $n$-bit string $r\in\{0,1\}^n$.
Second, for all $1\leq i\leq n$, if $r_i \neq p_i$ then the system
applies the classical \textsc{Not} operation, 
\begin{eqnarray*}
X & := & \left(\begin{array}{cc}0 & 1 \\ 1 & 0 \end{array}\right),
\end{eqnarray*}
to its bit of the entangled state $\ket{\psi_i}$.  If $r_i=p_i$,
then the $\ket{\psi_i}$ remains unchanged.  
Simultaneously, the sender applies $X$ to her part of $\ket{\psi_i}$
for all $i$ with $p_i= 1$, while she does not change the states $\ket{\psi_i}$ 
for which $p_i=0$.  (As a result, the states $\ket{\psi_i}$ have changed to
$(\ket{r_i\oplus p_i,p_i}+\ket{1\oplus r_i \oplus p_i,1\oplus p_i})/\sqrt{2}$.)
Third, the user measures her parts of the entangled states in the
computational basis $\{\ket{0},\ket{1}\}$, and sends these outcomes
$u_1,\dots,u_n$ to the system.  
The system measures his part of the entangled states in the computational 
basis as well, yielding the bits $s_1,\dots,s_n$.  Finally, the password is 
verified by the system who checks the identity 
$r_i = s_i\oplus u_i$ for all $i$.

First of all, it should be noted that the usage of the random bits $r_i$
in the above protocol is unnecessary.  The scheme works equally well
 if the system chooses $r$ to be the string $0\cdots 0$.
More important, though, is the fact that the same scheme can also be 
implemented when the parties substitute the entangled states $\ket{\psi_i}$
with  classically correlated states 
$\rho_i := \smfrac{1}{2}(\ketbra{00}{00}+\ketbra{11}{11})$.
Because the protocol uses only the classical \textsc{Not} operation 
and measurements in the computational basis, it is easily verified that 
this substitution gives a completely
classical process that does not differ from the just described procedure.
This shows that the security of the ``quantum password'' scheme does not 
rely on the quantum mechanical properties of the entangled states 
$\ket{\psi_i}$, but rather on the classical, shared randomness that 
these states exhibit.

In similar vein, one can recast the ``multiparty quantum
password'' scheme of \cite{mihara} into a classical procedure that 
uses classically correlated multipartite states. 
The essential ingredients of the scheme in \cite{mihara} are:
$m$-qubit states $\ket{\phi} := (\ket{0,\dots,0}+\ket{1,\dots,1})/\sqrt{2}$,
the \emph{Hadamard transform}
\begin{eqnarray*}
H & := & \frac{1}{\sqrt{2}}\left(\begin{array}{rr} 1& 1 \\ 1 & -1 \end{array}\right),
\end{eqnarray*}
the phase changing operation
\begin{eqnarray*}
Z & := & \left(\begin{array}{rr} 1& 0 \\ 0 & -1 \end{array}\right),
\end{eqnarray*}
and measurements in the computational basis $\{\ket{0},\ket{1}\}$. 
To see that this scheme is equivalent with a classical procedure, 
it is helpful to consider it in the \emph{Hadamard basis} with 
$\ket{\hat{0}} := H\ket{0} = (\ket{0}+\ket{1})/\sqrt{2}$ and 
$\ket{\hat{1}} := H\ket{1} = (\ket{0}-\ket{1})/\sqrt{2}$. 
Note that the phase changing $Z$ operation equals the classical 
\textsc{Not} operation $X$ in this basis: 
$H^\dagger\cdot Z\cdot H = X$,  which allows us to use the 
notation $\hat{X}=Z$.
Furthermore, we have for the initial entangled state, 
\begin{eqnarray*}
\ket{\psi} & = & 
\frac{1}{\sqrt{2^{m-1}}}\sum_{y_1\oplus\cdots\oplus y_m=0}{\ket{\hat{y}_1,\dots,\hat{y}_m}},
\end{eqnarray*}
which is a uniform superposition of all bit-strings $y\in\{0,1\}^m$ 
with even parity. 
In this alternative basis, the multipartite password scheme of \cite{mihara}
can be expressed as a protocol that uses the state 
$(\sum_{\oplus y=0}{\ket{\hat{y}_1,\dots,\hat{y}_m}})/\sqrt{2^{m-1}}$,
the \textsc{Not} transform $\hat{X}$ 
(with $\hat{X}:\ket{\hat{0}}\mapsto \ket{\hat{1}}$ and 
$\hat{X}:\ket{\hat{1}}\mapsto \ket{\hat{0}}$), and measurements
in the basis $\{\ket{\hat{0}},\ket{\hat{1}}\}$ 
(which equal the combination of the  Hadamard transform, followed 
by a standard measurement, which is used in the article).
As with the two-partite password scheme, we see by this limited
set of operations that the protocol relies only on the classical
correlations between the bits, only now it uses the Hadamard basis
instead of the computational basis.  As a result, we can again 
replace the entangled states $\ket{\psi}$ by classically 
correlated states
\begin{eqnarray*}
\rho & := & \frac{1}{2^{m-1}}\sum_{y_1\oplus\cdots\oplus y_m=0}
{\ketbra{\hat{y}_1,\dots,\hat{y}_m}{\hat{y}_1,\dots,\hat{y}_m}},
\end{eqnarray*}
without changing the cryptographic properties of the password scheme.

The ``identification'' and ``authentication schemes'' that are described 
in \cite{mihara} use states and operations that are identical 
to the ones of  the password scheme discussed above. 
Consequently, all of them can be re-expressed with classical 
correlated states, and none of them thus qualify to be called
``quantum schemes''.

In 1919, Gilbert Vernam patented\cite{vernam} the one-time pad 
encryption protocol where two parties use a shared, secret, 
random string $r_1,\dots,r_n$ to transmit a secret message 
$m_1,\dots,m_n$ by broadcasting publicly the scrambled string 
$m_1\oplus r_1,\dots,m_n\oplus r_n$.
Thirty years later, Claude Shannon proved the security of
this Vernam cipher in a mathematical rigorous way \cite{shannon}.
The recent article \cite{mihara} by Mihara described, 
essentially, this same classical procedure.
Both procedures require the parties to securely share 
correlated information---classical bits in the case  
of the Vernam cipher and, seemingly, quantum bits for
the implementation of Mihara's protocol.
It is true that the cited procedure of Lo and Chau 
\cite{lochau} gives an unconditionally secure way of 
distributing these correlations between the separated 
parties and it also true that this task does not seem to be 
possible classically.  But it would be incorrect to infer from 
this that the protocol of Mihara itself is quantum mechanical 
as well.
Rather, one has to conclude that it is only the distribution 
of correlations that requires the use of quantum mechanics,  
but not the usage of these correlations as described in 
Mihara's article.

\end{document}